\documentclass[aip,rsi,reprint,graphicx]{revtex4-1} 

\usepackage{graphicx}
\usepackage{epstopdf}
\usepackage{siunitx}

\begin{document}


\title{Note: Suppression of kHz-Frequency Switching Noise in Digital Micro-Mirror Devices} 



\author{Klaus Hueck}
\email[]{khueck@physik.uni-hamburg.de}
\affiliation{Institut fuer Laserphysik, University of Hamburg, 22761 Hamburg, Germany}
\author{Anton Mazurenko}
\affiliation{Department of Physics, Harvard University, Cambridge, MA 02138, USA}
\author{Niclas Luick}
\affiliation{Institut fuer Laserphysik, University of Hamburg, 22761 Hamburg, Germany}
\author{Thomas Lompe}
\affiliation{Institut fuer Laserphysik, University of Hamburg, 22761 Hamburg, Germany}
\author{Henning Moritz}
\affiliation{Institut fuer Laserphysik, University of Hamburg, 22761 Hamburg, Germany}


\date{\today}

\begin{abstract}
High resolution digital micro-mirror devices (DMD) make it possible to produce nearly arbitrary light fields with high accuracy, reproducibility and low optical aberrations. 
However, using these devices to trap and manipulate ultracold atomic systems for e.g. quantum simulation is often complicated by the presence of kHz-frequency switching noise. 
Here we demonstrate a simple hardware extension that solves this problem and makes it possible to produce truly static light fields.
This modification leads to a 47 fold increase in the time that we can hold ultracold $^6$Li atoms in a dipole potential created with the DMD. 
Finally, we provide reliable and user friendly APIs written in Matlab and Python to control the DMD.
\end{abstract}

\pacs{07.10.Cm, 07.50.Ek, 07.60.-j, 37.10.Gh, 42.79.-e, 42.79.Kr, 85.60.-q}

\maketitle 


The ability to generate arbitrary light fields is of interest to areas ranging from movie projection to quantum simulation with ultracold atoms. 
The digital micro-mirror devices (DMDs) developed by Texas Instruments for video projectors have made it possible to manipulate sub-micron sized systems of ultracold atoms with light potentials via holographic beam shaping \cite{Zupancic16} and direct imaging \cite{Liang2010,Gauthier16}. Highlights include the single site addressing of atoms in optical lattices to enable e.g. the observation of quantum mechanical random walks \cite{Greiner14} and the measurement of excitation spectra \cite{LiChung15}. 


The micro-mirror devices available on the market cover a vast range of features and characteristics, but most are oriented towards video projection applications.
Since most DMDs are not specifically designed to be used in a quantum optics context, nearly all devices lack essential features for the quantum optics end user e.g. the option to control the timing of the mirror switching.

\begin{figure}
	\includegraphics[width=1\linewidth]{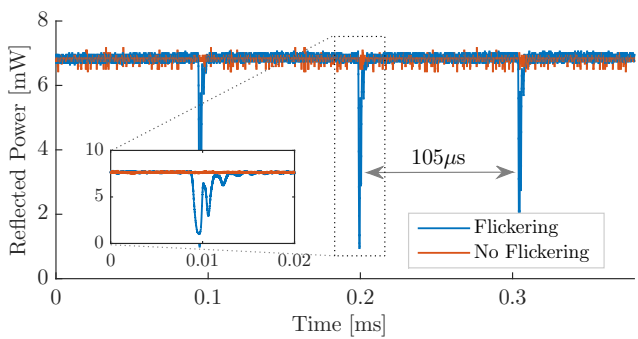}
	\caption{\label{fig:stolenClock} Reflection signal of the DMD with applied mirror clocking pulse (MPC) (blue line) and with the MCP pulled to ground (red line). In the latter case, the mirrors are not switched back to their \textit{flat}-state every $\SI{105}{\micro\second}$ and the generation of time stable light fields becomes possible. The inset shows one single switching event where additional spurious mirror ringing is visible.}
\end{figure}

DMDs typically consist of one to four million mirrors mounted on microscopic torsion springs above a static random access memory (SRAM) cell.
Each mirror can be pivoted by typically \ang{24} about a central support post via two electrodes which keep the mirror in either its \textit{on}- or its \textit{off}-position.
The endpoints of the states are fixed by landing pads which provide a well defined stop to the switched mirror.
The state of the mirror is governed by the state of the SRAM bit beneath it. 
When a mirror clocking pulse (MCP) addresses a particular pixel group, every pixel in this group is released, and subsequently settles to the state of the SRAM within \SIrange{3}{5}{\micro\second}.
The state of the mirrors when they are released during the MCP is termed the \textit{flat}-state. 

If the mirrors remain in the \textit{on}- or \textit{off}-state too long, the mirrors can either get stuck on the landing pads due to surface adhesion, or deform due to the applied stress.
For these reasons, DMD manufacturers usually implement a switching cycle which applies the MCP every clock cycle, even if the SRAM has not been updated.

This switching becomes apparent when a \SI{532}{\nano\meter} laser beam is reflected off the DMD with all pixels set to the \textit{on} state, and monitored by a photo-diode\footnote{Thorlabs PDA10A with \SI{150}{\mega\hertz} bandwidth.}.
The measured power vs. time is shown by the blue solid line in Fig.~\ref{fig:stolenClock}.
The device under test is a Texas Instruments (TI) DLP6500FYE\cite{Datasheet} implemented in the Lightcrafter EVM~6500 kit. The DMD features a resolution of 1080p, a maximum pattern rate of \SI{9.527}{\kilo\hertz} and a mirror pitch of \SI{7.6}{\micro\meter}.
Chips of this type are commonly used for quantum optics and quantum gas experiments as well as for spectral line-shaping of ultra-short light pulses.

\begin{figure}
	\includegraphics[width=1\linewidth]{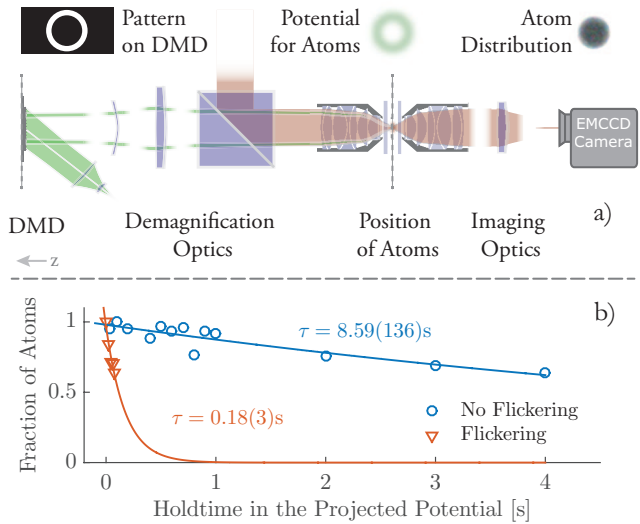}
	\caption{\label{fig:lifetime} Loss rate of ultra cold $^6$Li atoms. a) Sketch of the experiment. The Fermi gas with a Fermi energy of about $E_F/h=\SI{10}{\kilo\hertz}$ is trapped radially by a repulsive potential created by \SI{532}{\nano\meter} light reflected off a DMD. b) The lifetime is enhanced by a factor of 47 (blue circles) when the flicker of the DMD mirrors with \SI{9.5}{\kilo\hertz} is disabled as compared to the case where the mirrors flicker (red triangles).}
\end{figure}

As we show in Fig.~\ref{fig:lifetime}, the influence of this kHz-frequency mirror switching has detrimental effects in a quantum gas context: atoms trapped in a dipole trap formed by the reflection off the DMD surface are heated out of the trap due to the modulation of the dipole trap caused by the switching.
This is rarely a problem for heavy atoms such as Rb where the intrinsic time scales of the system are much slower than the switching frequency of the DMD\cite{Gauthier16}. However, it can lead to severe limitations when dealing with light atoms such as Li where time scales on the order of \SI{100}{\micro\second} are common.
In the following we show how a simple hardware modification makes it possible to disable this switching, increasing the trapping time by a factor of 47.

Two strategies are available to suppress the switching noise.
The first is to apply a modification to the firmware of the TI DLPC900 sequencer chip which controls the DMD or replace this chip with a field-programmable gate array (FPGA) with custom firmware.
The latter route has been taken by different companies\footnote{For example Vialux and BBS-Bildsysteme offer corresponding products.}.
It either requires access to the source code of the sequencer chip or considerable expertise in programming FPGAs.
Though possible, this route is challenging because the DLPC900 firmware is not freely available, and sufficiently high performance FPGAs are expensive and the implementation is technically challenging.
The alternative route, followed in this paper, is to access the desired signals at the hardware level in order to control the mirror switching.

\begin{figure}
 	\includegraphics[width=1\linewidth]{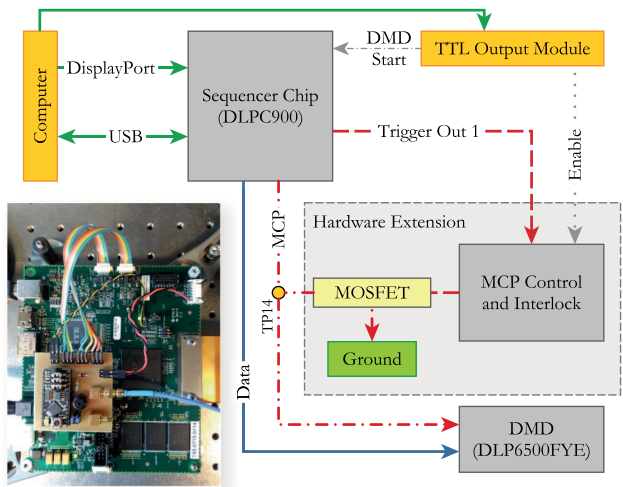}
 	\caption{\label{fig:circuit} Schematic of the hardware extension used to interrupt the MCP. The MCP can be pulled to ground via a MOSFET (red dashed-dotted line). The exact timing is determined via the ``Trigger Out 1'' signal generated by the DLPC900 (red dashed line). Whether or not the MCP should be pulled to ground can be set by a TTL signal coming e.g. from the control computer (gray dotted line). The inset shows the hardware implementation.}
\end{figure}

To do this, we interrupt the mirror clocking pulse (MCP) between the sequencer chip and the actual DMD, effectively freezing the DMD mirrors in their present position.
To load the next pattern from the sequencer chip to the DMD, the MCP is restored for a short time. 
This strategy is implemented using a simple external circuit interrupting the MCP, but only when instructed to do so by a readily available output-trigger of the sequencer.
This output-trigger indicates when new patterns are about to be loaded and when loading is finished.
A block diagram of the setup is shown in Fig.~\ref{fig:circuit}.

The external circuit interrupts the MCP by pulling it to ground when needed.
This is realized by a MOSFET with low on-resistance (e.g. BSS138CT-ND) connecting the strobe pin DADSTRB/AF5\cite{DatasheetDLPC900} of the DLPC900\footnote{This signal is available via test point TP14 on the control board.} to ground.
The DLP6500FYE loses the MCP and the mirrors freeze in their previous state (later on referred to as \textit{stop}).
When the ground connection is released, the MCP is routed towards the DMD again and the mirrors can switch freely (later on referred to as \textit{go}).
The sequencer's behavior is not affected by the fact that the DMD is missing its MCP during \textit{stop} phases.
In principle, shorting a signal to ground can pose an electrical danger to the integrated circuits.
Since the MCP trigger pulses are only nanoseconds long, the effective mean current shorted to ground is minute.

The control signal which sets the \textit{stop} and \textit{go} could be provided by any suitable logic signal. 
Here we use the ``Trigger Out 1'' signal of the sequencer chip: 
the DLPC900 allows to define its timing and polarity such that immediately after the sequencer displays a new pattern, the trigger is pulled high (\textit{stop}) and only returns to a low (\textit{go}) state \SI{20}{\micro\second} before a new pattern will be displayed. 
Therefore, the sequencer is capable of updating the pattern since during that time the MCP is routed towards the DMD.


As there is no data publicly available regarding the level of degradation of the DMD with exposure times longer than $\SI{4}{\second}$, an adjustable interlock circuit has been added to the system.
It monitors the timespan during which the state is set to \textit{stop}.
If this timespan is longer than a certain interval the MCP is routed back to the DMD for at least two MCP cycles. 
This precaution avoids freezing the mirrors for dangerously long times.
The scheme outlined above should be adaptable to DMDs other than the Lightcrafter EVM~6500 used here. 

The effectiveness of the presented hardware modification of the EVM~6500 board is shown by measuring the atom loss rate in a quantum gas experiment.
High loss rates are detrimental as they limit the trapping time and introduce decoherence into the system. 

The hardware modification has a huge impact on the ability to trap atoms, as we demonstrate by a loss rate measurement: A cloud of \num{20e3} $^6$Li atoms is radially confined in a box type potential. To do so, the DMD displaying a ring pattern is illuminated with \SI{532}{\nano\metre} light. The resulting intensity distribution is imaged onto the atoms, see upper panel of Fig.~\ref{fig:lifetime}, resulting in a repulsive potential. Axial trapping is provided by either a highly elliptical dipole trap\cite{Weimer2015} or an optical lattice potential.
Two measurements are performed to quantify the effect of the mirror flickering.
First, the mirrors are not frozen, the MCP is routed towards the DMD (red triangles in Fig.~\ref{fig:lifetime}) and atom numbers are measured via absorption imaging after a variable hold time. 
In the second measurement, the mirrors are frozen (blue circles in Fig.~\ref{fig:lifetime}).
The solid lines represent fits of the form \(f(t) = \exp(-t/\tau)\) to the two datasets.
The extracted lifetimes \(\tau\) are \SI{0.18+-0.03}{\second} and \SI{8.59+-1.36}{\second} for the flickering and the not flickering DMD respectively where the latter is limited by other factors.

This measurement demonstrates that eliminating the switching noise leads to a 47 fold increase in the lifetime of $^6$Li atoms trapped in a potential created with a DMD.

In this technical note we demonstrated the successful development and implementation of a simple circuit which makes it possible to suppress the switching noise in a DMD device.

The versatility of the EVM~6500 is further enhanced by releasing two open source implementations of control software packages written in Matlab and Python respectively\footnote{The source code is available via {https://github.com/deichrenner/DMDconnect} (Matlab) and {https://github.com/mazurenko/Lightcrafter6500DMDControl} (Python).} and allow for easy integration of the DMD into already existing environments.
This paves the way for the successful, yet easy integration of this device into experiments within the ultracold atoms and the broader quantum optics communities, whenever time stability of the projected patterns is of utmost priority and affordable devices with small pixel size and high resolution are required. 

Even though no detrimental effects have been observed on the DMD chip after daily use over the course of six months, we note that this hardware modification will void the warranty and can - according to the manufacturer - lead to reduced performance over extended timescales.\\

\section*{Supplementary Material}
See Supplementary Material at \url{ftp://ftp.aip.org/epaps/rev_sci_instrum/E-RSINAK-88-063701} for the full circuit diagram of the interlock device as well as the Gerber files and the microprocessor source code.
\\

This work has been financially supported by the European Union's Seventh Framework Programme ERC starting grant 335431 and by the DFG in the framework of SFB 925, GrK 1355 and the excellence cluster CUI.
A.M. acknowledges financial support from the NSF GRFP program, the ARO DARPA OLE Program, the AFOSR, the MURI, the ONR Defense University Research Instrumentation Program, and NSF.
We thank Jim MacArthur for helpful discussions and Kai Morgener and Jonas Siegl for their contributions to the experiment setup.



%
%

%

\bibliography{CTD}

\end{document}